\documentclass[12pt,dvips]{article}

\usepackage{amsmath,amssymb,exscale}
\usepackage{array,multicol}
\usepackage{afterpage,float,flafter}
\usepackage{epsfig,rotating,pifont}
\usepackage{cite}
\@addtoreset{equation}{section} \makeatother

\def\beq{\begin{eqnarray}}
\def\eeq{\end{eqnarray}}

\def\lsim{\mathrel{\rlap{\lower3pt\hbox{\hskip0pt$\sim$}}
    \raise1pt\hbox{$<$}}}         
\def\gsim{\mathrel{\rlap{\lower4pt\hbox{\hskip1pt$\sim$}}
    \raise1pt\hbox{$>$}}}         

\title{
\Large\textbf{Nature of Microscopic  Black Holes and Gravity  in Theories with Particle  Species}\\
\vspace*{.5cm}
\large \textbf{Gia Dvali \footnote{email: georgi.dvali@cern.ch, gd23@nyu.edu}}\\
\emph{CERN Theory Division, CH-1211, Geneva 23, Switzerland,} \\
 \emph{and CCPP,} \emph{Department of Physics, New York University}\\
\emph{4 Washington Place, New York, NY 10003}}

\date{}
\begin{document}
\maketitle \thispagestyle{empty} 

\begin{abstract}
Relying solely on unitarity and the consistency with large-distance black hole physics, we  
derive model-independent properties of the microscopic  black holes and of short-distance gravity 
in theories with $N$  particle species.  In this class of theories  black holes
can be as light as $M_{Planck}/\sqrt{N}$ and be produced in particle collisions above this energy.    
 We show, that the micro black holes must come in the same variety as 
the species do, although their label is not associated with any conserved charge measurable at large distances.  In contrast with big Schwarzschildian  ones, the evaporation 
of the smallest black holes is maximally undemocratic and is biased in favor of particular species.
  With an increasing mass the democracy characteristic to the usual macro black holes is gradually regained. The lowest possible mass above which black holes
become Einsteinian is $\sqrt{N} M_{Planck}$.
This fact uncovers the new fundamental scale
(below the quantum gravity scale)  
above which gravity changes {\it classically},   
and the properties of  
black holes and gravity are such as if some extra dimensions  open up, although no such input exists in the theory.  Our observations  indicate that  geometry of space may be an emergent  concept following from  large number of species and the consistency with macro black hole physics.     
 We apply our findings to the phenomenological properties of the micro  black holes that can be observed at LHC for large $N$.
Extrapolating our findings to  small $N$, one may ask whether the existence of quark and lepton flavors is already an evidence for emergent extra dimensions at short distances.   
\end{abstract}

\newpage
\renewcommand{\thepage}{\arabic{page}}
\setcounter{page}{1}

\section{Introduction}

  The main theoretical motivation for observing the physics beyond the  Standard Model around the 
  TeV scale is the Hierarchy Problem, an inexplicable smallness of  the weak interaction scale  relative to   the Planck mass, $M_P \sim 10^{19} $ GeV. The key problem is that the former scale is determined by the Higgs mass square, which  is {\it quadratically sensitive}  to an ultraviolet  (UV) cutoff of the theory.  In the absence of any new stabilizing physics at  intermediate scales, the natural cutoff of the theory is $M_P$, where the Einsteinian gravity gets strong and can no longer be neglected.  
  
    One approach to this problem\cite{ADD}  goes under the name of TeV scale quantum gravity, and 
 is based on the idea that UV divergences in the Higgs mass get cut off by gravity which gets strong 
 around the TeV energies.  The original framework realizing this proposal is based on large extra dimensions, in which there are $n$ compact extra dimensions of size $R$ 
to which gravity can propagate.  Then, there is the  following relation between $M_P$ and the fundamental high-dimensional Planck  mass ($M_*$), the cutoff of the low energy theory, 
\begin{equation}
\label{planck}
M_P^2 \, = \, M^2_*  \, (M_*R)^n \, .
\end{equation}   
This relation can be simply understood in terms of the high dimensional Gauss law, but for our purpose we shall focus on interpretation in terms of number of particle species.  For such interpretation it is enough to notice that the factor
$(M_*R)^n \equiv N_{KK}$ measures number of  Kaluza-Klein (KK) species with mass of order or below the cutoff. 
Thus, the relation (\ref{planck})  can be rewritten as 
\begin{equation}
\label{nbound}
M_P^2 \, = \, M^2_* \,  N_{KK}\, .
\end{equation}   
Thus,  in large extra dimensions the hierarchy between the $M_P$ and the gravity cutoff $M_*$ is set by  $\sqrt{N_{KK}}$.     This form of the relation is remarkable, since it allows to express the hierarchy 
between $M_P$ and the gravity cutoff in a geometry-independent way, solely in terms of the number of KK species.

 Relying on consistency with the large-distance black hole (BH) physics, 
it was understood recently\cite{largen,qg} that the latter  relation is much more general and holds true in the presence of 
 particle species irrespective of their geometric meaning. Namely, in a theory with  $N$ particle species,  
 both, the bound on their masses  and on the  gravitational cutoff of the theory is given by the scale 
 \begin{equation}
\label{nbound}
M_* \, \equiv \, {M_P \over \sqrt{N}} \, .
\end{equation}    
  Thus, any theory with large number of particle species lowers the cutoff, and can potentially address the hierarchy problem, irrespectively whether the species are 
  massive spin-2 states organized in the tower of KK states or are simply 
degenerated-in-mass
  lower spin particles.  For example,  $N$ exact copies of the Standard Model would do the same job. 

 The macroscopic BH physics indicates, that the masses of species 
are bounded by \cite{largen},  
 \begin{equation}
\label{massbound}
M_{species}  \, \lsim  \, {M_P \over \sqrt{N}} \, .
\end{equation} 
Consistency of the resulting BH physics with the well-known macroscopic BH properties, such as their thermal nature, also indicates the simultaneous lowering of the gravity cutoff\cite{qg}. These non-perturbative BH arguments are in excellent agreement 
with the perturbative
ones \cite{greg}, \cite{veneziano},\cite{qg} that signal the breakdown of perturbation theory at the scale (\ref{nbound}).  

The simultaneous lowering of the species masses and of the gravity cutoff also makes the whole picture consistent with the low energy effective field theory, since  the quadratic divergences in the scalar masses are cutoff by strong gravity at the scale $M_*$.  Below we shall be interested in this situation.

  Since gravity gets strong at $M_*$, the model-independent signature of this class of theories is the existence of micro black holes with the mass $M_{BH} \, \geq \, M_*$, and size $r_g\, \geq \, M_*$. In particular, such micro black holes can be produced in particle collisions at energies above $M_*$. 
    
     The purpose of this paper is to derive some generic properties of these micro black holes, without making any {\it a priory}  assumptions about the nature of the underlying microscopic gravity theory.  
 We shall rely  solely on the low energy consistency requirements, such as the 
unitarity and consistency of the low energy theory  with the well-understood large-distance black hole physics.  
      As we shall see, these arguments uncover the existence of the second length scale, larger than the fundamental length $M_*^{-1}$, at which gravity starts changing the regime {\it classically}. 
 Thus,  before becoming quantum in UV, gravity starts departing from the Newtonian gravity classically 
already at larger distances.
The absolute lower bound on the BH mass below which they can no longer 
be treated as normal Schwarzschild
BHs is given by 
\begin{equation}
\label{BHcros}
        M_{Schwarzschild~BH} \, \geq \, M_P\, \sqrt{N} \,.         
\end{equation}
 This bound is model independent, and in any theory with $N$ species, black holes 
lighter than $M_P \,\sqrt{N}$ must be non-Einsteinian. Notice that, in some theories the departure from the Einsteinian regime can start for even heavier BHs (see below), but never below the limiting mass (\ref{BHcros}).

  In the intermediate regime the properties of  the BH are rather peculiar. 
For example, their evaporation is undemocratic in species.  For the same mass BH, different species see
the different horizons, so that some do not see it at all.  This properties amplify with the decreasing mass. 
Heavier HBs on the contrary are more and more democratic, and closer to being Schwarzschildian.   
   By all accounts the  emerging gravitational dynamics  on intermediate scales is such, as if 
 some extra dimensions open up.  This fact may imply that geometry of the extra space is an emergent phenomenon, entirely forced upon us  by number of four-dimensional species, and consistency of the theory with a large distance BH physics.  In the other words species imply the change of  short distance  geometry!   Extrapolating our findings to moderate values of  $N$, one may ask whether  existence of 
 several quark-lepton species in the Standard Model can be regarded as the evidence for  such emerging dimensions in UV?

 \section{Power of Unitarity: Small Black Holes Carry ``Hair''} 
   
     In general the notion of ``black holeness''   in Einsteinian gravity is associated with the absence of ``hair''. 
  This association is based on the well-understood properties of the large  classical  black holes, that are known to satisfy no-hair theorems\cite{nohair}.  The absence of hair implies that large black holes  can carry information only about the 
  charges that can be measured at infinity, either classically or quantum mechanically. 
  
 Such are the electric or magnetic  type charges  associated with the massless gauge fields,  corresponding classical flux of which can be measured at infinity by the Gauss law,  or the charges associated  with  the Aharonov-Bohm type phases, which are undetectable classically but can be measured quantum-mechanically\cite{quantum}.
 
   As a result, in the absence of such detectable charges,  the Hawking evaporation process of a large
Schwarzschild black hole is completely democratic in species.   For example, in the presence 
of $N$ massless particle species, such a black hole will evaporate in each of these species  with the  same probability. 
  
     The nature of the micro black holes is dramatically different. As it was shown in  \cite{qg}, the
    decay  of a  lightest  micro BH  in a theory with $N$ particle species,  must be maximally undemocratic. This follows from unitarity.   
  Let us briefly repeat the argument here.   Consider a theory of $N$ particle species
 $\Phi_j$ with $j\, = \, 1,2,...N$, 
   that are  inter-coupled only through gravity. 
 Consider a smallest mass micro BH produced in the particle-anti-particle 
collision of $i$-th species.  
    Such a black hole can only be produced at energy $\sim M_*$, with a characteristic mass 
  $M_{BH} \sim M_*$.  Strictly speaking, the lightest  BH is still a quantum object,  but it is a strongly gravitating object at the distance of order its Compton wavelength 
and is  on the verge of becoming what one would call a
  quasi-classical BH.  Since both the center of mass energy and an impact parameter are continuous 
 variables, the transition from quantum to classical regime must be gradual.    
  Since the only scale in the problem is $M_*$,  production rate of such states is set by it  and we have 
  $\Gamma \sim M_*$.  Since the BH was produced in particle-anti-particle collision it is  
 neutral and  can decay into particle-anti-particle pairs of all the species.  By analogy with the classical Einsteinian BH one would expect this decay to be democratic
  in all the species.  
However,  this is impossible by unitarity.   By unitarity, the BH should be able to decay back into a pair of 
$i$-th species by the rate $\Gamma \sim M_*$, but then by the same unitarity,  it cannot decay into the other individual  species with  the same partial rates.  

 First, if a BH could decay democratically into each of  $N$ other species with partial rates exceeding  ${M_* \over N}$, the lifetime of  the BH would 
be less than its mass, which would indicate that it is not even a well defined state. In the other words, 
such a BH cannot exist. The latter conclusion would be impossible to reconcile with the fact
that gravity is getting strong at $M_*$ energies.  

 Secondly,  exchange of such a virtual BH would generate an effective operator 
 inducing the inter-species transitions 
 \begin{equation}
\label{transitions}
\Phi_i \,  + \, \bar{\Phi}_i  \, \rightarrow  \, \Phi_j \, + \, \bar{\Phi}_j,
\end{equation}
at the rate that would violate unitarity way below $M_*$ energies. 

For the consistent picture,  a small BH produced in the collision of $i$-th species should decay predominantly back into the same species, and only with very suppressed partial rates
\begin{equation}
\label{others} 
\Gamma_{BH \rightarrow  other~copy}  \lsim  {M_* /N}
\end{equation}
 into each of the other individual species.
 
   Notice, that for the small high-dimensional  BHs in large extra dimensions this property is explicit. Consider a  micro BH of mass $M_*$ produced in particle collision on a 3 brane where Standard Model species are localized.  From the point of view of the four-dimensional observer there are $N_{KK}$ additional species in form of the KK gravitons, however
   the evaporation rate into them and the Standard Model species is not democratic. 
 Decay  rate into each individual KK graviton is 
\begin{equation}
\label{rateKKind}
 \Gamma_{BH \rightarrow ~each~KK}\, \sim \,  {M_*\over N} \, , 
\end{equation}
which gives the total evaporation rate $\sim M_*$.  
 
 As a second example consider species that are localized on a distant brane. 
Consider a small BH produced in particle collision on ``our'' brane. If the high-dimensional gravitational radius 
is smaller than the inter-brane distance,  the  BH cannot evaporate into the species that are localized on the  distant brane.    

In both examples, the non-democracy has a clear geometric meaning, that can be understood as a 
consequence of locality in the extra space.  However, our discussion shows, that the geometric interpretation may be viewed as completely secondary, and non-democracy of the small BHs holds true 
in any theory with many species.

 Coming back to the four-dimensional setup with $N$ species,  we can repeat the same unitarity argument  for a BH produced in particle-anti-particle collision of each of the  $N$ species, and the situation is completely reciprocal.    
 Thus, we are lead to the conclusion that the smallest BHs are highly non-democratic, and have 
 a memory of belongness to a particular species.  Thus, small  BHs  come in $N$ distinct flavors.  In the other words, the 
 microscopic  BHs carry hair.  This hair is  not associated with any conserved  charge  measurable at large 
 distances either classically of quantum-mechanically,  and is microscopic in nature.  But its presence does not violate standard no hair theorems, as they are not  applicable.  Notice that micro BH are not Einsteinian,  for example their 
 mass to gravitational radius dependence is inevitably very different, from the Schwarzschildian ones.

   We shall now show that the above fact has the profound consequences for understanding the 
   nature of micro BHs.   We shall first derive the model-independent properties, and then apply to a concrete case when species are identical copies of each other  and are related by symmetry. 
  
   As we shall see, although no extra dimensions are postulated in our analysis, the unitarity 
 and other consistency arguments  tell us that  gravitational physics in the presence of species 
 is such, as if the space of species is a true dimension.   For example, the above mentioned
 memory of the micro BHs can be interpreted as a notion of locality in the imaginary space of species,  
 even though,  species were introduced as particle species in four dimensions.   This interesting
 fact 
 may indicate that the locality and geometry may be the emergent phenomena,  for accommodating consistency with BH physics in the presence of the 
multiple particle species.

    \section{Inevitability of the Larger Crossover Length Scale} 
    
    Non-democracy of the smallest BHs, inevitably leads to the existence of a second length  
    scale, which we shall denote by $R$. The key point is that $R$ is larger than  the fundamental  Planck length $M_*^{-1}$, and marks the crossover distance, beyond which the black holes become the normal Einsteinian  black holes. That is, the BHs larger than $R$ ``loose'' all the hair associated with the species flavor, and their evaporation 
 become democratic in all the species. 
     
 Indeed, we can prove inevitability of this second scale, by performing the following thought experiment. 
 Consider a micro BH produced in collision process of particles  and anti-particles of $i$-th species. 
 The smallest such  black holes will start getting created  for center of mass energies of order 
 $M_*$ and the impact parameter $\sim \, M_*^{-1}$.  At this point BH are still quantum and as argued above, by unitarity must  have a complete memory of the species of origin. Such BHs will quickly decay 
 into their own species, and with  $1/N$- suppressed probability to  most of the other individual species. 
 
  Now, rising the center of mass energy in this thought experiment, we produce heavier and bigger black holes, which gain more classicality, and correspondingly,  more  inter-species democracy in their evaporation process. This is inevitable for the consistent matching with the well known physics of large  Einsteinian BHs.
   
   Indeed, for example had we increased the center of mass energy up to a solar mass, while keeping the impact parameter small (at least smaller than 3 km, the solar Schwarzschild radius), we would start producing macroscopic  BHs  of the solar mass.  Obviously, the evaporation process of such a BH must  be completely democratic.  The evaporation rates  in all the thermally-available species (with masses below the Hawking temperature) are equal.  
   
   Thus, we see that by gradually increasing the mass of a micro BH, we make transition from the maximal
``favoritism" to the full democracy.  The critical size of the BH  horizon, at which this transition is complete,  
  we shall denote by $R$.  By default,  $R$ is related to the mass of the BH via the usual  Schwarzschild relation  $R \, = \, 2M_{BH} \, G_{Newton}$.  Obviously, in the same time $R$ marks the length scale beyond which,  for distances $r \, \gg \, R$ gravity is a normal Einsteinian gravity  with the usual $1/r$ Newtonian potential.  

 Thus, consistency of the black hole physics provides an important information about the nature of the 
 short distance gravity in theories with $N$-species.  Namely, we learn that gravity starts departing 
 from the Einsteinian (and Newtonian) gravity already at distances that are still  larger than the fundamental length $M_*^{-1}$.  The important fact is, that at intermediate distances 
\begin{equation}
 M_*^{-1} \, \ll \, r \,  \ll \, R \, , 
\end{equation}
 gravity is still classical, but the 
 micro BHs of  size (gravitational radius)  $r_g$ in the interval   $M_*^{-1} \, \ll \,  r_g \, \ll \, R$  
 are {\it not}  Einsteinian BH, and are  characterized  by  an  $r_g$-dependent  level of non-democracy with respect to the particle species.
   As said above, the black holes of sizes $r_g \, \sim \, M_*^{-1}$ and $r_g \, \gsim \, R$ are extreme cases, representing maximally undemocratic and maximally democratic objects. 
   
   Again, the distance $R$ in theories with extra dimensions has a very simple physical meaning. 
 It is the size of the compact extra dimensions.   But, what we are discovering is  that 
 existence of such crossover scale is a generic property of theories with many species.  
 This is remarkable, since at  no pint in our discussion we made any assumption or reference about the underlying  extra dimensional physics.  Yet  we are finding an outcome  which normally would  be 
 interpreted as a signal of an underlying geometry.   
  
 Notice that in principle the distance scale $R$ can parametrically be 
close to $M_*^{-1}$. In extra dimensional example,
this is the case when 
number of extra dimensions is large, since 
$R \rightarrow M_*^{-1}$ as $n\rightarrow \infty$.
 This possible proximity is however an illusion, since the separation of the relevant energy scales is controlled by $N$.  
Indeed, to probe the length scales between $M_*^{-1}$ and $R$, the BH mass must interpolate within the interval $M_*^{-1} \, <\, 
M_{BH} \, < \, M_P^2R$. Even for the lowest possible value $R\sim M_*^{-1}$ 
this corresponds to the $N$-fold change in the energy scale. In the other words, the 
absolute lower bound on the Einsteinian BH mass is given by (\ref{BHcros}). 
The shortest possible mass interval in which BH are non-Einsteinian, therefore is
\begin{equation}
\label{interval}
{M_P \over \sqrt{N}} ~ \leq \, M_{Non-Enisteinian~BH}  \,\leq  ~ M_P\sqrt{N} \, .
\end{equation}
  At this point of the discussion we know the properties of the BHs, such as their masses and decay rates into various species,  only at the endpoints of the non-Einsteinian 
interval.   We know that for $r_g \sim M_*^{-1}$ and  $r_g \sim R$, the masses of the black holes are  $M_{BH} \, \sim \, M_*^{-1}$ and 
 $M_{BH} \, \sim \, RM_P^2$ respectively.   However,  we do not know what is the dependence of 
 $M_{BH}$ and of the BH evaporation rate on  $r_g$   within the interval. 
  
   We now wish to investigate these questions. 
  First, notice that, since in the interval of interest the properties of the BHs are not democratic with respect 
  to all the species, the notions such as BH horizon require certain care.  In the other words, horizon 
  with respect to some species, may not be seen as a horizon for the others.  This simply follows from the 
  fact that intermediate size BH are not evaporating universally to all the thermally-available species. 
  By consistency,  the species that are not produced in the evaporation of a given BH, cannot see the horizon of this BH. Such species can go through the given BH untrapped.  
   
    A simple explicit example is provided by species that live on different branes separated by a large extra dimension. Obviously, horizon of a BH created by species on a given brane, will not be seen  as a horizon  by species on the neighboring brane,  if the high-dimensional Schwarzschild  radius of a BH is smaller than the inter-brane separation.   Even if  four-dimensional coordinates  of particles on the second  brane are within the 
  four-dimensional projection of the BH gravitational radius, they cannot be captured by the BH.   
 Correspondingly the BH will not be able to evaporate into these species.  
 What we are discovering, however, is that irrespectively whether we wish for it, some sort of extra dimensional behavior emerges whenever we talk about many species.

  What the above fact tells us is that at the intermediate distances the gravitational force between two particles  depends on which species they come from.     
  The two particles (or the localized sources) of the same species with the center of mass energy $\sim M_*$  should gravitate strongly  (with $\sim M_*$ gravitational potential),   when separated by the distances $\sim M_*^{-1}$. This  is not necessarily true when the two particles belong to the different species.  Again, this is obvious from the extra dimensional example, since the two point-like sources of mass $M_*$ localized on the two largely-separated  3-branes, will gravitate weakly even if their four-dimensional coordinates coincide within the radius $M_*^{-1}$.  
  
   Therefore,  let us first analyze gravity of sources and probes belonging to the  same $i$-th flavor of species. 
 Consider a Newtonian gravitational interaction between the two point-like sources of masses $M_i$
 and $m_i$
 separated by a distance $r$, where index $i$ indicates that sources are made out of $i$-th species. Since we are making no assumption about the existence of any extra dimensions, 
$r$ will always refer to the distance in our three-dimensional space.
  When $r$ changes  from $R$ to  $M_*^{-1}$, this Newtonian interaction interpolates  between the usual Newtonian force and the strong gravity. This interpolation can be parameterized by the following gravitational potential (below everywhere we shall ignore numbers of order one)  
 \begin{equation}
\label{v}
V(r) \, = \, {M_im_i \over M_*^2} {1 \over  r  \, \nu (M_*r)_{ii}} \, ,   
\end{equation} 
 where $\nu(M_*r)_{ii}$ is  some smooth function,  such that
 \begin{equation}
\label{nu}
\nu \sim 1~~{\rm for} ~~ r\sim M_*^{-1},~~ {\rm and}~~ 
\nu \simeq  {M_P^2 \over M_*^2} ~~{\rm for}~~ r \, > \, R \, . 
\end{equation} 
The indexes $ii$ indicate that gravitational force is measured between the two sources belonging 
to the $i$-th species. 
Since at intermediate distances gravity is weakly coupled,  gravitational radii of the sources can be determined in Newtonian approximation, from the condition that gravitational potential becomes order one.

Thus,  the effective gravitational radius ($r_g^{ii}$)  of the source of mass $M_{BHi}$
seen by the particles of the same $i$-th species can be estimated from the equation
  \begin{equation}
\label{v}
1\, = \,  {M_{BHi} \over M_*^2} {1 \over  r_g^{ii} \, \nu (M_*r_g^{ii})_{ii}} \, .   
\end{equation} 
We shall refer to this radius as the self-species horizon. 

Crudely speaking,   $r_g^{ii}$ marks a distance from the source at which the Newtonian escape velocity for a massive particle of the same species 
becomes order one, and thus $r_g^{ii}$ can be thought as the BH horizon for the same species.
 However, as was argued above, $r_g^{ii}$  is not necessarily a horizon for all the other species. 
 Only the species that can be produced in the evaporation process of a given BH, can see its horizon.  For any BH belonging to $i$-th site,  
the number of such species is a function of the black hole mass, or equivalently of its self-species 
gravitational radius.  
$\mathcal{N}(M_*r_g^{ii})_{i}$.  Again, the index $i$ labels the site of origin in the
space of species. 
The boundary properties of this function are similar to (\ref{nu}) 
\begin{equation}
\label{n}
\mathcal{N}(1)_i \sim 1, ~~~ \mathcal{N}(M_*R)_i =  {M_P^2 \over M_*^2}
\end{equation} 


 In general, if we are interested in gravity between the two sources of masses $M_i$ and $M_j$  belonging to the two different $i$-th and $j$-th species , we should consider the function $\nu(M_*r)_{ij}$.   For $i\neq j$ the only 
model independent boundary condition is $\nu(MR)_{ij} = M_P^2/M_*^2 = N$, which follows from the definition of $R$.  The necessary 
condition for the particles of $j$-th species to be produced in the evaporation process of the BH 
of mass $M_{BHi}$  made out of  $i$-th species, is that  $j$ particles see the horizon of the $i$-th BH. 
The corresponding horizon we shall call $r_g^{ij}$.  Generalizing notion of  
$r_g^{ii}$,  we can try to estimate this quantity again in Newtonian approximation as the radius where 
Newtonian gravity of the $i$-th BH  exerted  on the $j$-th probe source becomes strong.  
That is,  
\begin{equation}
\label{evap}
{M_{BHi} \over M_*^2} {1\over r_g^{ij} \, \nu(M_*r_g^{ij})_{ij}} \, = \, 1 \, .
\end{equation} 
Then $r_g^{ij}$ marks the gravitational radius (horizon) of $i$-th BH with respect to the particles from the 
$j$-th species.   Then, the function $\mathcal{N}_i$ counts the solutions of (\ref{evap}).  

  The species-dependence of functions $\nu_{ij}$ have a well defined meaning  in large extra dimensional example.  There the species localized at different extra coordinates will exert different force on each other.   Correspondingly, they will 
see different four-dimensionally-projected horizons of a given high-dimensional BH.  
  This is defined by the metric in the extra dimensional space, and by  sites of localization of  species. 
Small BH cannot ``reach out'' to the species localized on the distant branes. 
   Again, what we are finding is that some notion of metric emerges in the ``space of species'', irrespectively of any input. 
  
   This phenomenon  is remarkable because of the following reason.  Of course, since species are the orthogonal states,  formally one can always  think of them as being eigenstates of either
  coordinate or  momentum operators in some mathematical space.  That is, the label $j$ can formally 
  correspond to some extra coordinate $y_j$ or a momentum  $p_j \, \equiv \, -i \, {\partial \over \partial_{y_{j}}}$ .   This is a triviality.  The non-trivial part however is that this space inevitably modifies laws of gravity, and thus is not just a mathematical formality. 
      
 Coming back to the definition of  $r_g^{ij}$, the following subtlety arises. Is could happen that 
 for a given species $r_g^{ij}$ is defined in a probabilistic sense.  This will happen if $i$ and 
 $j$ are not in the same (coordinate or momentum) representations in the space of species. 
  For example, say $\Phi_i$ is an eigenvalue of $y_i$, whereas $\Phi_j$ is an eigenvalue of 
  $p_j$.  Of course, one can try to go in the same representation by taking a new superpositions of species
  \begin{equation}
\label{superp}
 \sum_j \, c_j \Phi_j, 
\end{equation}  
 but the new states will no longer be mass eigenstates  in general. 
 So to work with mass eigenstates, besides $r_{g}^{ij}$-s we have to introduce the probabilities of 
 capture  $P_{ij}$ within the radius $r_g^{ij}$. By unitarity this is 
related to the emission probability.     
  The black hole evaporation rate then will be given by 
\begin{equation}
\label{masschange}
{d M_{BHi} \over d t}  \, = \, -  \, \sum_j \, (r_g^{ij})^{-2}\, P_{ij}  \, . 
\end{equation} 
The summation is over thermally available states only (or else one has to explicitly include 
Boltzmann-suppression factors).
This expression looks singular for $r_g^{ij} \, = \, 0$, but it is not, since in this case $P_{ij} \, = \, 0$. 
In the other words, the particles that cannot see the horizon, cannot be produced. 
 The non-zero quantities $(r_g^{ij})^{-2}\, P_{ij} $ cannot exceed a given fixed value,
and for a static black hole presumably must be equal by thermodynamics arguments. 
 Notice that the above expression correctly reproduces the evaporation of the high-dimensional BH 
 in KK theories.  Applied to evaporation of a BH  of gravitational radius 
$r_g \,  \ll \,R$ in  $n$ extra dimensions
 of size $R$,  
 we get  for each KK graviton 
 \begin{equation}
\label{example}
(r_g^{ij})^{-2}\, P_{ij} \, = \, r_g^{-2}  \left ({r_g \over R}\right )^n.
\end{equation} 
 Summation  up to the thermally available species gives an extra factor  $\left ({r_g \over R}\right )^{-n}$, which gives the correct evaporation rate 
 \begin{equation}
\label{masschange}
{d M_{BHi} \over d t}  \, = \, -  \, r_g^{-2}\, . 
\end{equation} 

  We shall consider now the class of theories in which the form of these functions is highly restrictive. 
 This is the situation when $N$ species are exact copies of each other, related by an exact permutation symmetry.

 \section{Micro Black Holes  in N-copies of Species} 
 
 We shall now consider a special case in which the species are identical copies of each other and are related by some exact permutation symmetry.  Namely we require that physics is identical as seen from each copy.  
  In this case, by the symmetry  the quantities  such as $r_g^{ii}$,  $\nu_{ii}$ and $\mathcal{N}_i$ are independent of $i$. Moreover for $N \gg 1$,  if we require some analog 
 of flatness in the space of species,  then functions 
 $\nu$ and 
 $\mathcal{N}$ can be approximated by  
   \begin{equation}
\label{nuandn}
\nu \, = \, \mathcal{N} \, = \, \left ( {r \over R} \right )^n {M_P^2 \over M_*^2} \, , 
\end{equation}
where $n$ is an arbitrary number (not necessarily an integer).
The boundary condition $\mathcal{N}(1) = 1$ fixes $R \, = \, (M_P^2/M_*^2)^{{1\over n}} M_*^{-1}$, and thus, 
\begin{equation}
\label{nuandn1}
\nu \, = \, \mathcal{N} \, =\, (r M_*)^{n}\,. 
\end{equation}
In this case gravity is such as if  the species are equally spaced in $n$ extra dimensions.

 Let us compute the lifetime of a neutral BH of mass $M_{BH}$ produced in the collision of particle and antiparticle of a given species.  Gravitational radius of such a BH can be determined from (\ref{v}).  Taking into the account (\ref{nuandn1}),  we get 
 \begin{equation}
\label{radius}
r_g \, = \, M_*^{-1} \, \left ( {M_{BH} \over M_*} \right )^{{1\over 1 + n}}
\end{equation} 
 Not surprisingly, the relation between $r_g$ and $M_{BH}$ are such as if the space has $n$ flat extra dimensions (with $n$ not necessarily an integer), although no extra dimensions have ever been postulated.   The change of mass due to Hawking evaporation for any  BH of Hawking temperature 
$T_H$, 
 is given by 
\begin{equation}
\label{masschange}
{d M_{BH} \over d t}  \, = \, -  \, T_H^4(M_{BH}) r_g^2 \, \mathcal{N}(M_{BH}),
\end{equation} 
 where both the Hawking temperature and the number of available species are expressed as functions of 
 $M_{BH}$.  Since  $T_{H} \, = \, r_g^{-1}$, using (\ref{nuandn1}), we get
 \begin{equation}
\label{masschange2}
{dM_{BH} \over dt} \, = \, M_*^2 \, \left ( {M_* \over M_{BH}} \right )^{{2-n \over 1+n}}, 
\end{equation}
which for the BH lifetime gives 
\begin{equation}
\label{lifetime}
\tau_{BH} \, \simeq \, M_*^{-1} \left ({M_{BH} \over M_*}\right ) ^{{3 \over n +1}} { n+1 \over 3}.   
\end{equation}  
For $n=0$ this reproduces the lifetime of an usual Schwarzschild BH, as it should.  Notice, that this lifetime is by a factor $\left ({M_{BH} \over M_*}\right )^{{n \over n+1}}$ shorter than the lifetime
of a BH in $n$ flat extra dimensions with a single 3-brane, and is approximately the lifetime of a BH in $n$ flat extra dimensions in which 
there are $N$ 4-dimensional species localized at
uniformly  distributed sites (3-branes).  
Again, none of this was input in our discussion.  Yet,  we see that requiring exact symmetry between the species,  effectively leads to the behavior as if  the species are localized on $N$ uniformly distributed branes in $n$ flat  large extra dimensions.\footnote{It was noted by Michele Redi, that had we gone beyond the requirement of 
identical physics as seen from each site, and assumed a stronger symmetry 
under the full permutation group, this would correspond to a maximal departure from 
the flatness in the species space.}
 
   It is also clear that throughout  the evaporation process of a heavy BH  ($M_{BH} \, \gg \, M_*$) produced by particles of $i$-th copy,   only a fraction 
   \begin{equation}
\label{fraction}
{E_{BH \rightarrow i-{\rm th~copy}} \over  E_{BH \rightarrow  {\rm all~copies}}} \sim \left ({M_* \over M_{BH}}\right )^{{n\over n+1}}
\end{equation}
of the total energy will be released in $i$-th species. The rest will be distributed over  $\mathcal{N} \, = \, \left ({M_{BH} \over M_*}\right ) ^{{n \over n+1}}$ extra copies.

\section{$10^{32}$ Standard Model Copies} 

 We  are now ready to apply our findings to the situation of $N\, = \, 10^{32}$ Stanard Model copies. 
 With this choice the fundamental Planck mass becomes  $M_* \equiv M_P/\sqrt{N} \sim$TeV, as needed for the solution of the hierarchy problem.  
 
 Consider now an experiment with particle collision performed in our copy of the Standard Model (which we can call simply the 
 Standard Model).  By default, around TeV energies and TeV$^{-1}$ impact radius, the micro BHs start being created.  According to our findings, the smallest mass BHs will decay back to the Standard Model states, and with an extremely suppressed probability into the other copies.  It will be virtually impossible to distinguish these lightest BH-s from some new unstable particles, which promptly decay back into the 
 standard model states, plus some new invisible states.  
    However,  once the center of mass energy 
 climbs above the fundamental scale, the  BH properties will quickly unfold.   Both  the lifetime, and the 
 fraction of the missing energy will rise according to (\ref{lifetime}) and (\ref{fraction}) respectively. 
 In the same time, the amplitudes must soften, with the typical momenta of the scattering products 
 being cutoff by the inverse BH size $r_g^{-1}$ given by (\ref{radius}).  This is a characteristic BH behavior. Notice, that the expected softening of the scattering 
products is a general qualitative property for the processes with quasi-classical black hole formation regardless of number of species (see \cite{soft}, and references therein).

  By lifting the requirement of exact symmetry, one effectively introduces ``warping'' in the space of species. This may affect form of the functions $\nu$ and $\mathcal{N}$,  but for small deformations the  
qualitative picture should remain as discussed above. 

 What about the constraints on parameter $n$?  As we saw,  just postulating  $N$ exact copies of 
 species, gives rise to the gravitational behavior characteristic of extra dimensions.  So it is tempting to identify  $n$ with (generically fractional) number of extra dimensions. However, irrespectively whether 
 we wish to do so, the important physics is that $N$ species do imply new classical gravitational physics 
 at the length scale $R$  that is related to $n$ the way radius of the extra space would be.  This is enough for putting constraints of $n$.   The  immediate phenomenological constraint $n > 2$, comes from the absence of new gravitation forces at the distances down to fraction of a millimeter \cite{adelberger}. 
  However,  unlike the standard extra dimensional case, we have no argument against non-integer $n$. 
  So in principle $n$ could be any fractional (or even irrational)  number exceeding $2$ or so. 

\section{Discussions and Outlook}

   We have seen that the low energy consistency requirements, such as unitarity
   and consistency of theory with large-distance BH physics, are  powerful tools  in understanding the general properties of microscopic gravity and BHs in the presence of 
$N$ particle species.  
  
  Our analysis uncovers the existence of the second scale, below the gravitational 
cutoff of the theory, above which gravity starts departing from the Einsteinian (Newtonian) gravity {\it classically}, even before reaching the quantum regime.   
Correspondingly, the model-independent lower bound on the mass of Einsteinian 
BHs is set by (\ref{BHcros}). 

 In the window between the two scales, BH gradually depart from the known classical properties, such as 
 democracy in the evaporation process and the absence of hair.  Smallest BHs therefore are maximally undemocratic and come in the same variety as the species do. 
 
  The overall picture is such as if some new dimensions open up, although nothing like this was input in the theory.  This fact may indicate that geometry is a secondary concept, which emerges in order to 
  accommodate  consistency of microscopic physics with the low energy unitarity and macro BH physics, 
  in the presence of species.   The natural question then is whether our findings can be extrapolated for 
  moderately small values of $N$, such as the number of quark lepton flavors  in the standard model.  
If yes,  then the existence of flavors may be a direct evidence for emerging new dimensions in UV. 
Of course, if the number of species is limited by the Standard Model alone, corresponding length scale will be small, but  this is conceptually unimportant.

 On the other hand if the solution of the hierarchy problem is related to the number of species,
 LHC should observe BH production in particle collisions, irrespectively of the precise origin of the species. 
 The versions of the theory in which species  are organized in a symmetric way are especially predictive. 
 The two opposite extremes are,  flat large extra dimensions with $N_{KK} = 10^{32}$ KK states, 
  and  $N=10^{32}$ copies of the standard model.  
    The properties of the BHs, such as their evaporation rate and the  fraction of missing energy, are different in the two cases.

  {\bf Acknowledgments}

We thank Oriol Pujolas, Michele Redi, Sergey Sibiryakov and Sergey Solodukhin for discussions and comments. This work  is supported in part  by David and Lucile  Packard Foundation Fellowship for  Science and Engineering, and by NSF grant PHY-0245068.



\begin{thebibliography}{99}

\bibitem{ADD} 
N.~Arkani-Hamed,  S.~Dimopoulos and G.~Dvali, {\it Phys. Lett.}  {\bf B 429} (1998) 263, hep-ph/9803315.  {\it Phys. Rev.} {\bf D 59} (1999) 086004, hep-ph/9807344.

I.~Anatoniadis,  N.~Arkani-Hamed, S.~Dimopoulos and  G.~Dvali, {\it Phys. Lett.} {\bf 436} (1998), 
hep-th/9804398. 
\bibitem{largen} 
 G.~Dvali, arXiv:0706.2050 [hep-th]

 For generalization to deSitter space, see, 
G.~Dvali  and D.~Lust, arXiv:0801.1287 [hep-th]
 
\bibitem{qg} 
 G.~Dvali and M. Redi, arXiv: 0710.4344 [hep-th].

\bibitem{greg}
G.~Dvali and G.~Gabadadze,
{\it Phys. Rev.}  {\bf D 63}, 065007 (2001) [hep-th/0008054].


\bibitem{veneziano}
  G.~Veneziano,
  JHEP {\bf 0206}, 051 (2002)
  [arXiv:hep-th/0110129].


\bibitem{nohair} 
 W.~Israel,  {\it Phys. Rev.} {\bf 164} (1967) 1776;  {\it Commun. Math. Phys.}
{\bf 8}, (1968) 245;  

B.~ Carter, {\it Phys. Rev. Lett.}  {\bf 26}  (1971)  331.

J.~Hartle, {\it Phys. Rev.}  {\bf D 3} (1971) 2938.


  J.~Bekenstein, {\it Phys. Rev. }\  {\bf D 5}, 1239 (1972);
  {\it Phys.\ Rev.}\  {\bf  D 5},  (1972) 2403; {\it Phys. Rev. Lett.} {\bf 28} (1972) 452. 
  
  C.~Teitelboim, {\it Phys. Rev.} {\bf D 5} (1972) 294.  
  

 
  
   \bibitem{quantum}
 BH quantum hair under discrete gauge symmetries was first discussed in,
 
  L.M.~Krauss and F.~Wilczek,  {\it Phys. Rev. Lett.} {\bf 62} (1989) 1221.
  J.~Preskill and L.M.~Krauss, {\it Nucl. Phys.} {\bf B341} (1990) 50. 
 S.~Coleman, J.~Preskill and F.~Wilczek, {\it Phys. Rev. Lett.} {\bf 67} (1991) 1975.
  
 Axionic quantum hair of BHs was discovered in,

M.J.~Bowick, S.B.~Giddings, J.A.~Harvey, G.T.~Horowitz, A.~Strominger,
{\it Phys.Rev.Lett.} {\bf 61} (1988) 2823.
T.J.~Allen, M.J.~Bowick, A.~Lahiri,
{\it Phys.Lett.} {\bf B237} (1990)  47

  Massive spin-2 quantum hair of BHs was found in,

  G.~Dvali,  {\it Phys. Rev.}  {\bf D74} (2006) 044013, hep-th/0605295; 
  hep-th/0607144.

\bibitem{soft} 
G. Veneziano, ~JHEP 0411:001,2004, hep-th/0410166.


\bibitem{adelberger}

D.J. Kapner, T.S. Cook, E.G. Adelberger, J.H. Gundlach, B.R. Heckel, C.D. Hoyle, H.E. Swanson,
{\it Phys. Rev. Lett} {\bf 98} (2007) 021101, hep-ph/0611184



%

\end{thebibliography}
\end{document}